\documentclass[10pt,conference]{IEEEtran}
\usepackage{cite}
\usepackage{amsmath,amssymb,amsfonts}
\usepackage{algorithmic}
\usepackage{graphicx}
\usepackage{textcomp}
\usepackage[dvipsnames, svgnames, x11names]{xcolor}
\usepackage[hyphens]{url}
\usepackage{fancyhdr}
\usepackage{hyperref}
\usepackage{pifont}
\usepackage{listings}

\newcommand{\hpcayear}{2025}

\newcommand{\hpcasubmissionnumber}{78}
\title{SoMa: Identifying, Exploring, and Understanding the DRAM Communication Scheduling Space \\for DNN Accelerators}

\def\hpcacameraready{} 

\newcommand\hpcaauthors{Jingwei Cai$^{\dagger}$, Xuan Wang$^{\ddagger\S}$, Mingyu Gao$^{\dagger\P\spadesuit}$, Sen Peng$^{\ddagger\S}$, Zijian Zhu$^{\dagger}$, Yuchen Wei$^{\dagger}$, Zuotong Wu$^{\ddagger\S}$, and Kaisheng Ma$^{\dagger\ast}$}
\newcommand\hpcaaffiliation{Tsinghua University$^{\dagger}$, Xi'an Jiaotong University$^{\ddagger}$, IIISCT$^{\S}$, Shanghai AI Laboratory$^{\P}$ Shanghai Qi Zhi Institute$^{\spadesuit}$ \\Corresponding Author$^{\ast}$}
\newcommand\hpcaemail{\{caijw21,gaomy,zhuzj23,weiyc22,kaisheng\}@tsinghua.edu.cn, \{xuanwang,3123353083\}@stu.xjtu.edu.cn}

\author{
  \ifdefined\hpcacameraready
    \IEEEauthorblockN{\hpcaauthors{}}
      \IEEEauthorblockA{
        \hpcaaffiliation{} \\
        \hpcaemail{}
      }
  \else
    \IEEEauthorblockN{\normalsize{HPCA \hpcayear{} Submission
      \textbf{\#\hpcasubmissionnumber{}}} \\
      \IEEEauthorblockA{
        Confidential Draft \\
        Do NOT Distribute!!
      }
    }
  \fi 
}

\fancypagestyle{camerareadyfirstpage}{%
  \fancyhead{}
  
  \fancyhead[C]{
    \ifdefined\aeopen
    \parbox[][12mm][t]{13.5cm}{\hpcayear{} IEEE International Symposium on High-Performance Computer Architecture (HPCA)}    
    \else
      \ifdefined\aereviewed
      \parbox[][12mm][t]{13.5cm}{\hpcayear{} IEEE International Symposium on High-Performance Computer Architecture (HPCA)}
      \else
      \ifdefined\aereproduced
      \parbox[][12mm][t]{13.5cm}{\hpcayear{} IEEE International Symposium on High-Performance Computer Architecture (HPCA)}
      \else
      \parbox[][0mm][t]{13.5cm}{\hpcayear{} IEEE International Symposium on High-Performance Computer Architecture (HPCA)}
    \fi 
    \fi 
    \fi 
    \ifdefined\aeopen 
      \includegraphics[width=12mm,height=12mm]{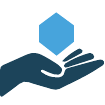}
    \fi 
    \ifdefined\aereviewed
      \includegraphics[width=12mm,height=12mm]{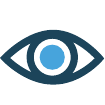}
    \fi 
    \ifdefined\aereproduced
      \includegraphics[width=12mm,height=12mm]{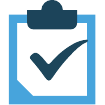}
    \fi
  }
  \fancyfoot[C]{}
}
\begin{document}
\maketitle

\ifdefined\hpcacameraready 
  \thispagestyle{camerareadyfirstpage}
  \pagestyle{empty}
\else
  \thispagestyle{plain}
  \pagestyle{plain}
\fi

\newcommand{\hpcaheight}{0mm}
\ifdefined\eaopen
\renewcommand{\hpcaheight}{12mm}
\fi

\newcommand{\mingyu}[1]{\textcolor{purple}{[~MINGYU:~#1~]}}
\newcommand{\experiment}[1]{\textcolor{Black}{#1}}
\newcommand{\SA}[1]{\textcolor{Black}{#1}}
\newcommand{\tool}[1]{\textcolor{Black}{#1}}
\newcommand{\valid}[1]{\textcolor{Brown}{#1}}
\newcommand{\open}[1]{\textcolor{Black}{#1}}
\newcommand{\hardwareconfig}[1]{\textcolor{Black}{#1}}
\newcommand{\hardwaredes}[1]{\textcolor{olive}{#1}}
\newcommand{\motivation}[1]{\textcolor{DarkGrey}{#1}}
\newcommand{\batch}[1]{\textcolor{Tan}{#1}}
\newcommand{\cut}[1]{\textcolor{Orange}{#1}}
\newcommand{\reuse}[1]{\textcolor{Black}{#1}}
\newcommand{\baseline}[1]{\textcolor{Black}{#1}}

\begin{abstract}

Modern Deep Neural Network (DNN) accelerators are equipped with increasingly larger on-chip buffers to provide more opportunities to alleviate the increasingly severe DRAM bandwidth pressure. However, most existing research on buffer utilization still primarily focuses on single-layer dataflow scheduling optimization. As buffers grow large enough to accommodate most single-layer weights in most networks, the impact of single-layer dataflow optimization on DRAM communication diminishes significantly. Therefore, developing new paradigms that fuse multiple layers to fully leverage the increasingly abundant on-chip buffer resources to reduce DRAM accesses has become particularly important, yet remains an open challenge.

To address this challenge, we first identify the optimization opportunities in DRAM communication scheduling by analyzing the drawbacks of existing works on the layer fusion paradigm and recognizing the vast optimization potential in scheduling the timing of data prefetching from and storing to DRAM.
To fully exploit these optimization opportunities, we develop a Tensor-centric Notation and its corresponding parsing method to represent different DRAM communication scheduling schemes and depict the overall space of DRAM communication scheduling. Then, to thoroughly and efficiently explore the space of DRAM communication scheduling for diverse accelerators and workloads, we develop an end-to-end scheduling framework, SoMa, which has already been developed into a compiler for our commercial accelerator product. Compared with the state-of-the-art (SOTA) Cocco framework, SoMa achieves, on average, a 2.11$\times$ performance improvement and a 37.3\% reduction in energy cost simultaneously. Then, we leverage SoMa to study optimizations for LLM, perform design space exploration (DSE), and analyze the DRAM communication scheduling space through a practical example, yielding some interesting insights. Moreover, SoMa has been open-sourced at https://github.com/SET-Scheduling-Project/SoMa-HPCA2025.

\end{abstract}
 
\section{Introduction}

\noindent In order to process a variety of tasks with better performance and accuracy, DNNs are rapidly becoming more complex and heavy~\cite{Resnet,transformer,bert,vision_transformer,gpt4}. Accelerators~\cite{TPU,tpuv2,TPUlesson,graphcore,huawei,cerebras2} with more computing units, larger buffers, and higher memory bandwidth have been developed to accelerate these DNN workloads. 

However, under modern semiconductor process, the rate of increase in DRAM bandwidth lags behind the rate of transistor density growth, which is a long-standing issue~\cite{nature_BW,memorywall}. This disparity is even more pronounced in DNN accelerators~\cite{nature_BW,chimera}, which are more specialized and have a larger proportion of dedicated computing elements than traditional chips like CPU. Thus, DRAM communication is increasingly becoming a performance bottleneck in DNN computation~\cite{nature_BW,chimera}. To mitigate this bottleneck, accelerators are equipped with increasingly larger on-chip buffers~\cite{tenstorrent_micro_report,tpuv2,TPUlesson,graphcore,huawei,cerebras2}, which provide opportunities to optimize DRAM communication by exploiting reuse opportunities in DNNs. 


Several works leverage buffer resources to reduce DRAM accesses under the paradigm of ``layer fusion''~\cite{fused-layer,full-hd,mei2023defines,kao2023flat,chimera,cocco,SET,cai2024gemini}. This approach involves buffering the feature maps (fmaps) produced by earlier layers on-chip, allowing subsequent consuming layers to read them directly. This strategy avoids the overhead of first writing back to DRAM and then reading from it, thereby reducing DRAM access costs. This optimization paradigm holds immense potential; for instance, Cocco~\cite{cocco} achieved performance improvements ranging from 1.89\% to 50.33\% by merely exploring the option of which layers to fuse. Beyond this, there are numerous dimensions worth exploring under this paradigm, such as execution order and execution granularity. However, like Cocco, most existing studies~\cite{fused-layer,mei2023defines} have only focused on a small portion of this optimization space. Therefore, we believe that the complex optimization dimensions within the layer-fusion paradigm have yet to be clearly delineated or defined, much less thoroughly explored and understood in the context of the entire layer-fusion optimization space.

While reducing DRAM access is an important approach for optimizing DRAM communication, we identify another optimization approach that has been largely overlooked in the field of DNN scheduling: prefetching and delayed storing, i.e., adjusting the timing of fetching/storing data from/to DRAM to be sometime earlier/later.
We focus on this approach and believe it has potential based on an insightful observation: in modern DNN networks, the ratio of DRAM bandwidth demand to computing demand varies significantly across different layers (see Fig.~\ref{figure:ratio}). After layer fusion, the overall ratio of DRAM bandwidth demand to computing demand for different tiles (computing unit) varies even more (see Fig.~\ref{figure:ratio}). This observation indicates that with the application of layer fusion, DRAM bandwidth usage during the entire computing process becomes very uneven—sometimes leading to congestion due to high demand and sometimes causing bandwidth resource waste due to low demand. This motivates us to apply prefetching-and-delayed-storing techniques to alleviate the uneven DRAM communication load. However, choosing the appropriate timing for prefetching and storing is a non-trivial problem. Clearly defining, thoroughly exploring, and understanding this paradigm is a significant challenge.

``Layer Fusion'' and ``Prefetching and Delayed Storing'' each have their own optimization spaces, but they are not independent and are intricately coupled. This is reflected in the following points: 1) both paradigms trade buffer usage for DRAM communication optimization, leading to competition for buffer usage, and 2) layer fusion affects the types and quantities of data that need to communicate with (i.e., prefetch from and store to) DRAM. Therefore, we define the complex space formed by these two paradigms as the DRAM Communication Scheduling Space. 


After identifying the challenges and optimization potential of DRAM communication scheduling, we make the following contributions to identify, explore, and understand DRAM Communication Scheduling Space.

We first introduce a Tensor-centric Notation with two categories and a total of six attributes to encode the scheduling schemes in the DRAM Communication Scheduling Space; then we show how to parse each encoded scheme into actual hardware behaviors. Based on this notation, we define and illustrate the DRAM Communication Scheduling Space and the complex trade-offs behind it. 
Existing works can be described using our notation, representing only a small subset within the space we have defined. To the best of our knowledge, this work is \textbf{\textit{the first}} to comprehensively define and analyze the DRAM Communication Scheduling Space.

To thoroughly and structurally explore the DRAM Communication Scheduling Space defined by the Tensor-centric notation, we develop an end-to-end framework, SoMa, which employs a Buffer Allocator, a two-stage simulated annealing (SA) exploration engine, and an accurate simulator to conduct a structured and efficient exploration of the space.
We have successfully built a complete compilation flow based on SoMa, from model input to instruction generation, for an accelerator approaching mass production. 

Then, we conduct extensive experiments on different workloads (including CNNs and LLMs), hardware configurations, and batch sizes, demonstrating an average of a 2.11$\times$ performance improvement and a 37.3\% reduction in energy costs compared to the SOTA framework, Cocco. We analyzed the experimental performance of LLMs and uncovered several intriguing phenomena: (1): For the decode stage, the optimization potential of DRAM scheduling is minimal due to its extremely low compute density, which imposes a pure DRAM bandwidth demand on the accelerator. (2): For the decode stage, increasing the batch size does not consistently improve computational utilization. As the batch size grows, the increasing size of the KV cache becomes comparable to that of the weights, diminishing the benefits of further increases in batch size for improving compute density. In addition, we use SoMa to explore and analyze the architectural design space, gaining some interesting insights. For example, with small batch sizes, DRAM bandwidth plays an irreplaceable role. However, with SoMa, the importance of the buffer becomes increasingly prominent as the batch size increases. Moreover, we present a practical execution graph comparison between Cocco and SoMa to enhance understanding of the trade-offs underlying the DRAM Communication Scheduling Space.

\begin{figure}[!t]
    \centering
    \includegraphics[width=2.8in]{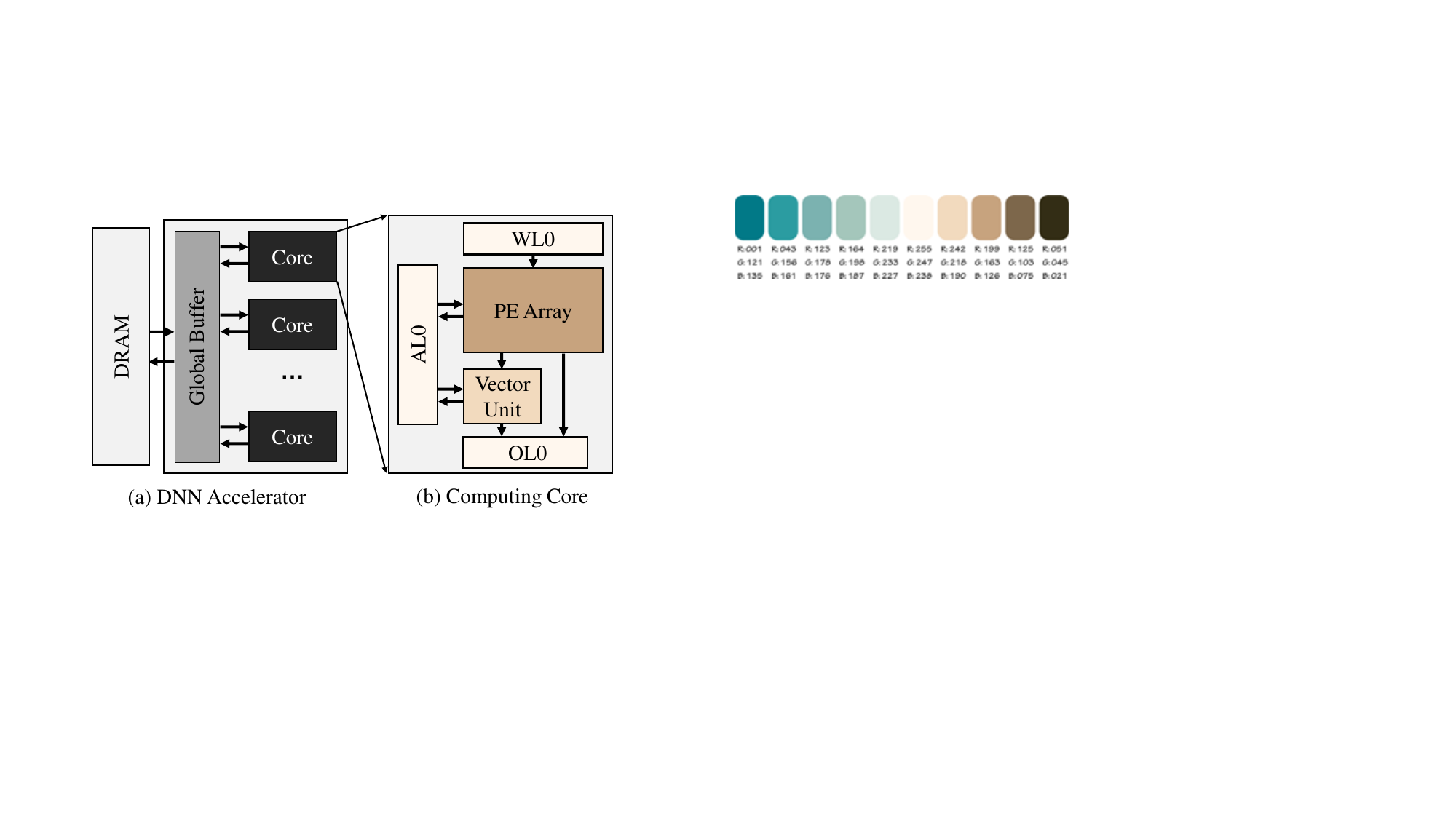}
    \caption{DNN Accelerator Template}
    \label{figure:hardware}
     \vspace{-5mm}
\end{figure}

\section{Hardware Basis}\label{ch2_hardware_baseline}

\noindent In this section, we introduce a generic large-scale DNN accelerator template (Fig.~\ref{figure:hardware}) and the corresponding abstract instruction system adopted in this paper, which represents many mainstream commercial DNN accelerators~\cite{TPU,tpuv2,TPUlesson,meta_mtia,huawei,hanguang,tesla}.

As shown in Fig.~\ref{figure:hardware}(a), the template primarily consists of DRAM, a Global Buffer (GBUF), and several cores. The GBUF is shared among all cores. As shown in Fig.~\ref{figure:hardware}(b), each core has private small buffers/register files ($WL0, AL0, OL0$) for rapid access by computing units. The PE Array is used for computing GEMM/Conv operations, and the Vector Unit is designed for computing other vector/scalar operations, such as element-wise addition, pooling, layer normalization, etc.

Although specific instructions vary significantly among these accelerators, they still share apparent common patterns. Based on these common patterns, we abstract three instructions: load, store, and compute. The ``load'' and ``store'' instructions refer to moving data from DRAM to the GBUF and from the GBUF to DRAM, respectively. The ``compute'' instruction refers to the operations performed on a tensor/vector. In accelerators, a tensor is often divided into smaller tensors, which are sequentially processed by the core group. Each small tensor is further split into smaller sub-tensors for parallel processing by the cores within the core group. The specific operations involved include loading ifmaps and weights from GBUF into the local buffer of each core, performing the computations, and then writing the computed ofmaps back to the GBUF. Since these instructions typically occur in sets and are synchronized, and as this study focuses on optimizing DRAM communication, we abstract them into a single ``compute'' instruction. In our DRAM-COMPUTE diagram (e.g., Fig.~\ref{figure:example} Bottom), ``load \& store'' instructions and ``compute'' instructions can be respectively represented by tensor blocks in the DRAM row and the COMPUTE row. The start and end of any instruction can serve as markers for the beginning of another instruction (Fig.~\ref{figure:example} Right).


\section{Identify DRAM Communication Optimization Opportunities}
\noindent The GBUF plays a crucial role in optimizing DRAM communication within DNN accelerators, and research on how to use it to optimize DRAM communication has been a hot topic since the advent of DNN accelerators~\cite{diannao,dadiannao,pudiannao,eyeriss,TPU}. However, most current studies focus on optimizing dataflow for small-scale DNN accelerators, specifically on how to tile a single layer into small parts and adjust their computing order to optimize DRAM communication~\cite{timeloop,understand,interstellar,baton,magnet,confuciuxtushka,CoSA,hasco,tenet,mindmapping}.

As DNN accelerators have evolved, architects have equipped them with increasingly larger GBUFs, with capacities reaching tens or even hundreds of megabytes~\cite{TPUlesson,huawei,meta_mtia,groq,graphcore,hanguang}. Such large buffers can accommodate most individual layers of most networks, significantly reducing the effects of optimizing single-layer dataflow for DRAM communication. Therefore, developing new techniques to fully leverage the rapidly growing on-chip buffer resources to optimize the increasingly bottlenecked DRAM bandwidth is crucial and offers significant opportunities. Next, we will analyze the challenges and optimization opportunities in DRAM communication scheduling.

\begin{figure}[!t]
    \centering
    \includegraphics[width=3in]{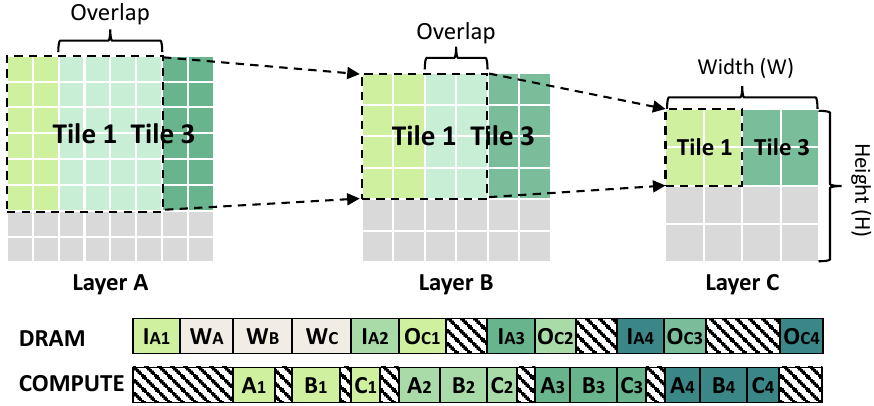}
    \caption{A Practical Layer-fusion Group ($LG$) Example}
    \label{figure:motivation}
     \vspace{-4mm}
\end{figure}

\subsection{Layer Fusion}\label{ch2:fusion}
\noindent Layer fusion is an important paradigm for using buffers to optimize DRAM communication~\cite{fused-layer,cocco,mei2023defines,EfficientSchedulingliuleibo}. Fig.~\ref{figure:motivation} illustrates the computation and DRAM access behavior of a simple three-layer network under layer fusion, where multiple fused layers are computed in a fine-grained manner sequentially, relying on on-chip buffers for switching fmaps. The optimization dimensions within this paradigm are numerous and complex, but have not been clearly depicted and analyzed. A straightforward optimization dimension is which layers to fuse, which affects both DRAM access savings and buffer occupancy. This is the primary focus of most existing studies~\cite{cocco,EfficientSchedulingliuleibo}. Another evident optimization dimension is the computing granularity of each fused layer group, i.e., the size of each computing tile (Fig.~\ref{figure:motivation}). This affects buffer occupancy, halo overlap overhead, and the effectiveness of intra-core optimizations (detailed analysis is in Sec.~\ref{ch3.1.1:LFA}). DeFENIS~\cite{mei2023defines} has analyzed this dimension but has not jointly explored the above two dimensions. Most other studies address this dimension using heuristic rules~\cite{cocco,EfficientSchedulingliuleibo}, overlooking optimization opportunities within this space. We can see that even these two dimensions lack systematic joint exploration, let alone additional dimensions (introduced in Sec.~\ref{ch3.1.1:LFA}) that have been overlooked by existing works.



\begin{figure}[!t]
    \centering
    \includegraphics[width=3in]{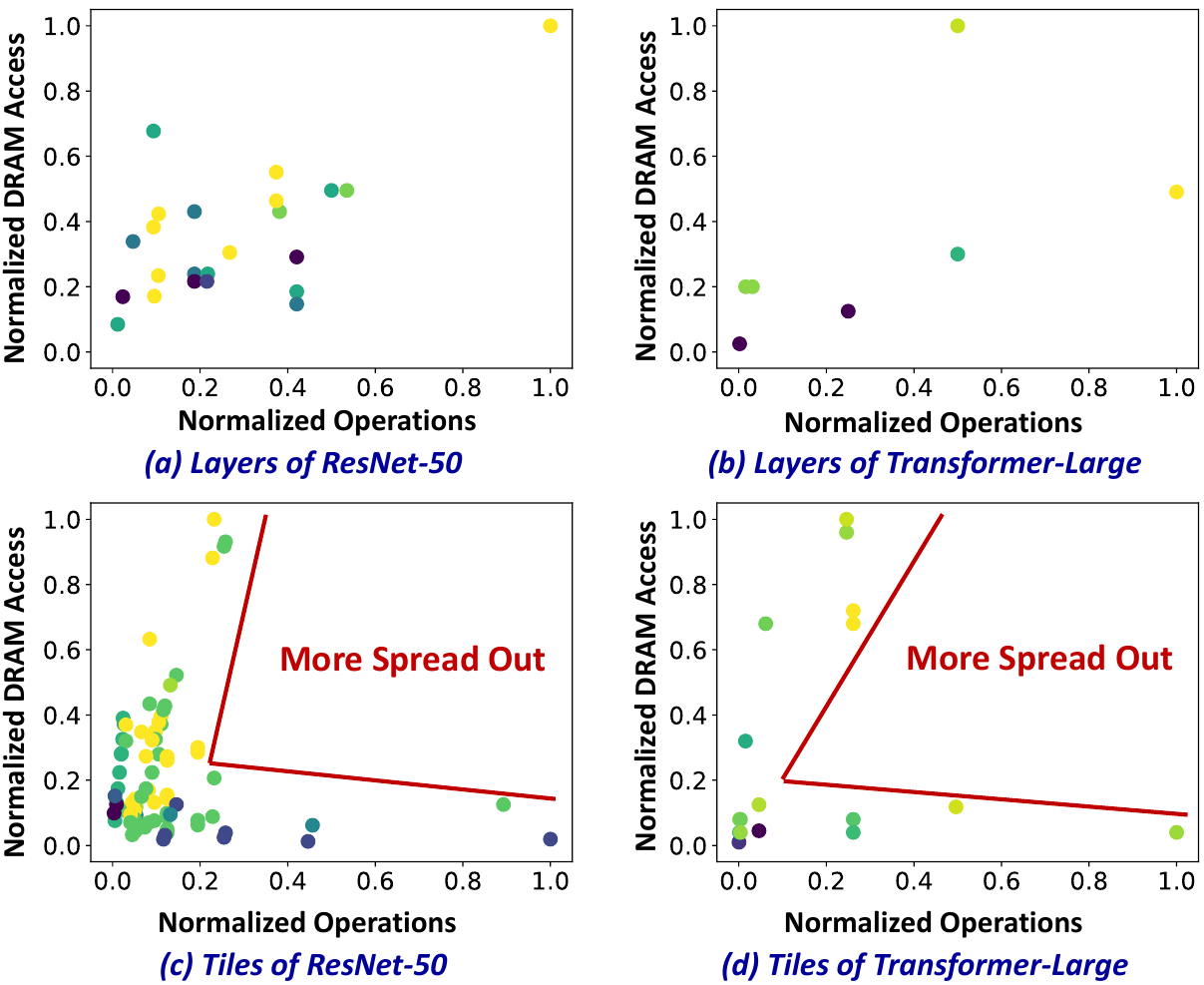}
    \caption{(a) and (b) show the normalized DRAM access and the normalized operation number for each layer in ResNet-50 and Transformer-Large, respectively (each point represents a layer). (c) and (d)  show the normalized DRAM access and the normalized operation number for each smallest computing unit (Tile) of ResNet-50 and Transformer-Large, respectively, scheduled using the SOTA Cocco Framework (each point represents a Tile). \textbf{\textit{The darker the color, the more identical overlapped points there are}}. 
    The normalization method involves dividing the value of each point by the maximum value among all points (DRAM access and operations are independently normalized). We use the default edge accelerator and batch size 1, as introduced in Sec.~\ref{sec:setup}}
    \label{figure:ratio}
    \vspace{-4mm}
\end{figure}
\begin{figure*}[!t]
    \centering
    \includegraphics[width=7.2in]{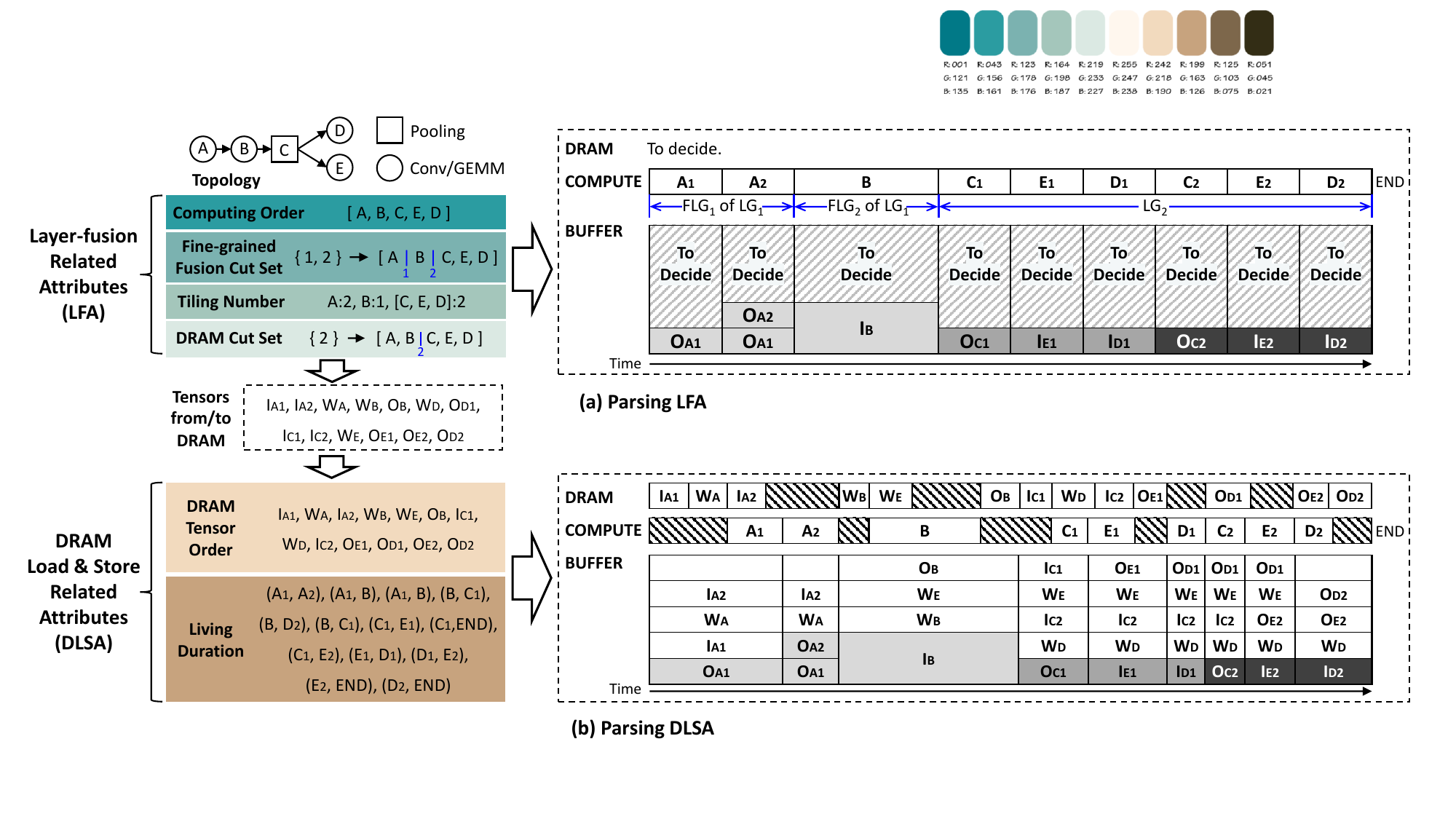}
    \caption{Parsing an Example Encode of a Five-Layer Network into Actual Scheduling Schemes. The DRAM, COMPUTE, and BUFFER in the right part show the DRAM access, workload computation, and buffer usage, respectively. We use $M_{i}$ to represent the $i$th tile of layer $M$, and $I/O_{Mi}$ to represent the ifmaps and ofmaps of $M_{i}$. $W_{M}$ represent the weights of layer $M$. In the BUFFER, blocks with the same background color represent the same data.}
    \label{figure:example}
     \vspace{-4mm}
\end{figure*}
\subsection{Prefetching and Delayed Storing}\label{ch2_prefetch_delay}
\noindent Besides layer fusion, we have identified another optimization opportunity overlooked by existing literature: prefetching and delayed storing. We argue this paradigm also has great potential based on the following insightful observation.


As shown in Fig.~\ref{figure:ratio}(a) and (b), in modern DNNs, the DRAM access demand and computing requirements vary significantly across different layers. Unfortunately, after layer fusion, the variance in the memory access to computation ratio for each tile becomes even more pronounced (see Fig.~\ref{figure:ratio}(c) and (d)), which are more spread out along the axes compared to their counterpart in Fig.~\ref{figure:ratio}(a) and (b). This indicates a larger number of DRAM-access-intensive and compute-intensive tiles, with fewer tiles having balanced demands. Specifically, this is because, within a fused layer group, the first tile of each layer with weights needs to load weights (e.g., $A_{1}, B_{1}, C_{1}$ in Fig.~\ref{figure:motivation}), which often results in high DRAM demand, corresponding to the points near the Y-axis in Fig.~\ref{figure:ratio}(c) and (d). The subsequent tiles do not need to load weights (e.g., $A_{2}, B_{2}, C_{2}$ in Fig.~\ref{figure:motivation}).
Additionally, many fmaps' DRAM access requirements are eliminated as a result of fusion (e.g., the ofmaps of $A_{i}$ and $B_{i}$). Consequently, many tiles even have no DRAM access demand (e.g., $B_{2}, B_{3}, B_{4}$), corresponding to the points near the X-axis in Fig.~\ref{figure:ratio}(c) and (d).

This severe imbalance between DRAM access and compute demands makes the overlap between DRAM access and compute under the traditional double-buffer strategy (prefetching data in the previous tile and storing data in the next tile) insufficient, especially in the context of layer fusion. For example, Fig.~\ref{figure:motivation} shows the DRAM communication under the traditional double-buffer strategy. As seen, there is a waste of DRAM bandwidth, and the computing resources are severely stalled. To better demonstrate the severity and prevalence of this challenge, we analyze the workloads, batch sizes, and platforms in Sec.~\ref{sec:setup}. With the SOTA Cocco Scheduling Strategy, the DRAM and computation utilization rates for the case in Fig.~\ref{figure:motivation}(c) are 52.69\% and 62.64\%, respectively, while in Fig.~\ref{figure:motivation}(d) they are 72.45\% and 45.84\%. These utilization are defined as the ratio of the sum of all DRAM tensors/computing tiles time to the total runtime. This result indicates that neither resource is fully utilized, leaving significant opportunities for overlap.

Based on the above observation and analysis, we find that controlling the timing of data prefetching and storing to utilize idle DRAM bandwidth and alleviate peak-time pressure has great potential. For example, suppose that there are two consecutive identical Layer-fusion Groups ($LG$ and $LG'$) (Fig.~\ref{figure:motivation} shows $LG$, and $LG'$ is not shown). By prefetching $LG'$'s $W_{A}', W_{B}'$, and $W_{C}'$ during the DRAM idle time corresponding to $B_{3}, B_{4}$, and $C_{4}$ in $LG$, the computing stall at the start of $LG'$ can be resolved. Additionally, starting to load $I_{A3}$ and $I_{A4}$ in $LG$ earlier can also erase the stalls before $A_{3}$ and $A_{4}$ in $LG$. However, achieving precise and automatic control over prefetching and delayed storing is non-trivial. Clearly defining, thoroughly exploring, and understanding this paradigm is a significant challenge.





\subsection{Combine Them Together}\label{ch3_3:combine_them_together}

\noindent ``Layer Fusion'' and ``Prefetching and Delayed Storing'' are two optimization paradigms with complex interrelationships, which lie in two aspects. The first aspect lies in the fact that both optimization paradigms inherently trade buffer usage for DRAM communication optimization, leading to competition for buffer resources between the two. Additionally, the layer fusion choice affects DRAM communication requirements, which in turn impacts the optimization space of prefetching and delayed storing. 
Therefore, these two paradigms exhibit complex interactions within the DRAM Communication Scheduling Space. Efficiently and structurally exploring this space presents a major challenge. Addressing this challenge and gaining a deep understanding of these interactions form the key objectives of this paper.

\section{Tensor-centric Notation} 

\subsection{Encoding Format and Parsing Methods}\label{ch3_1:notation}
\noindent In this section, through a practical example of a five-layer network ($A$ to $E$) shown in Fig.~\ref{figure:example}, we explain how the proposed Tensor-centric Notation translates layers into fine-grained tensor computations, DRAM accesses, and buffer usage, as well as the trade-offs associated with different encoding choices.


As shown in Fig.~\ref{figure:example}, the notation has six attributes, which can be divided into two categories: Layer-Fusion-related Attributes (LFA) and DRAM-Load-and-Store-related Attributes (DLSA). Consequently, the overall parsing process is also divided into two stages. The first stage involves parsing the LFA to determine: 1) the computing granularity and order, as well as the buffer occupancy of tensors reused on-chip, and 2) all tensors requiring DRAM interaction. The second stage involves parsing the DLSA to derive the specific timing and buffer occupancy details for all tensors that must be loaded/stored from/to DRAM.

\subsubsection{Layer-Fusion-related Attributes}\label{ch3.1.1:LFA}

\noindent LFA includes \textbf{\textit{Computing Order, Fine-grained Layer-fusion Cut (FLC), Tiling Number,}} and \textbf{\textit{DRAM Cut}}, which we will discuss in sequence.


The first attribute, Computing Order, arranges these layers into a serial sequence, representing their coarse-grained execution order. Thus, a valid Computing Order cannot have any dependency that goes from right to left, as this would result in a scenario where the data needed by a layer calculated earlier is not yet computed. We can see an example of the Computing Order in Fig.~\ref{figure:example}; swapping the order of $D$ and $E$ remains valid, but swapping $A$ and $B$ does not.


The second attribute, Fine-grained Layer-fusion Cut (FLC), records the cut locations, which cut layer sequences into Fine-grained Layer-fusion Groups (FLGs). All FLCs together constitute the FLC Set ($\{1,2\}$ in Fig.~\ref{figure:example}). Each FLG possesses an attribute, Tiling Number (the third attribute), which determines the computing granularity of the FLG. Layers within an FLG are processed continuously at the granularity of the tile rather than completely finishing the first layer before starting the second, thereby saving on-chip buffer overhead. For the example in the left top of Fig.~\ref{figure:example}, the layers are cut into three FLGs ($[A], [B], [C,E,D]$) with Tiling Numbers $2,1,2$, respectively. Within the FLG $[C, D, E]$, each layer is divided into two tiles, which are then processed sequentially. 
Given the Tiling Number, we use a heuristic strategy to partition each layer into computing tiles along the multiple dimensions.
Specifically, it prioritizes tiling the batch dimension since it does not produce halo overlap costs, followed by the height and width of the ofmaps, keeping them as equal as possible to reduce overlap. The reason for not splitting the channel dimension is that splitting the channel would prevent the next layer from accessing all channels, making it impossible to fuse more than two layers. For example, in Fig.\ref{figure:motivation}, with a batch size of 1 and a Tiling Number of 4, we split the height and width dimensions each by 2 (the channel dimension is not shown). It is worth emphasizing that for intermediate layers involving operations like convolutions or poolings, which produce halo overlaps, the size of each tile may be larger than $1/Tile$ of the size of the fmaps (e.g., $Layer A$ and $B$ in Fig.~\ref{figure:motivation}). The specific method for determining each tile's size of intermediate layers considering halo overlap influence has already been proposed in Cocco~\cite{cocco} and DeFENIS~\cite{mei2023defines}, so we directly adopt their methods. The trade-off associated with the Tiling Number primarily concerns the balance between buffer usage, halo overlap overhead, and computing efficiency within the core array. The finer-grained tiles imply reduced buffer usage, but if they involve layers that produce overlaps, they may incur more extra computation and memory accesses. Additionally, the computing efficiency within the core array may decrease, as smaller tiles imply reduced on-chip reuse opportunities~\cite{interstellar}.

By parsing the above three attributes, we can derive the entire computation sequence (the COMPUTE row in Fig.~\ref{figure:example}). 

We define certain specific FLCs as DRAM Cuts, which cut layer sequences Layer-fusion Groups (LGs). Dependencies between different LGs require data to be sent to DRAM and then loaded back for computation. All DRAM Cuts together constitute the DRAM Cut Set ($\{2\}$ in Fig.~\ref{figure:example}). For example, $C$ depends on $B$, and there is a DRAM cut between them. Therefore, $C$ can only load ifmaps from DRAM after $B$ has sent its ofmaps to DRAM. Thus, all the requests for each tile's interaction with DRAM can be determined as follows: if a layer has weights, it has a weight-load request. For example, all layers except $C$ (pooling layer has no weights) require their weights to be loaded from DRAM (represented as $W_{M}$ in the left part of Fig.~\ref{figure:example}). If a layer has forward dependencies that span different LGs (or its input is the overall network input), then the ifmaps of its related tiles need to be loaded from DRAM ($I_{Mi}$ in the left part of Fig.~\ref{figure:example}). If it has backward dependencies that span different LGs (or its output is the overall network output), then the ofmaps of its related tiles need to be written back to DRAM ($O_{Mi}$ in the left part of Fig.~\ref{figure:example}). 
The trade-off involved in DRAM Cuts primarily concerns the balance between buffer requirements and the volume of DRAM access. Generally speaking, the more fused layers there are (the fewer the DRAM Cuts), the lower the number of DRAM accesses will be, but the demand for buffer capacity will increase.

The remaining fmaps corresponding to the dependencies that do not cross DRAM cut can be directly reused on-chip. The buffer occupancy duration for such data ranges from the production of the ofmaps tile to the consumption of the tile. For the example in Fig.~\ref{figure:example}(a), there is an FLC between $A$ and $B$, thus the ofmaps of $A$ are stored on-chip. Fig.~\ref{figure:example}(a) only shows the buffer allocation for on-chip data transfers, while all tensors related to DRAM communication are not displayed (to determine in the DRAM Load \& Store Phase).

Since the DRAM Cut Set is a subset of the FLC Set, each LG contains one or more FLGs. When data dependencies exist between different FLGs, the producing layers in the former FLG need to aggregate the ofmaps from different tiles of the layers before they can be used by the consuming layers in the latter FLG. Once the computation for an FLG is completed, the weights of the layers in the FLG can be released, freeing up some buffer space. For the example in Fig.~\ref{figure:example}, the weights of $A$ can be released after all its tiles are computed. $B$ must wait until $A$ to finish computing all its ofmaps before it can start its computation. The above analysis demonstrates that the FLC primarily involves a trade-off in buffer occupation. Specifically, the FLC can free up some buffer space occupied by weights at the cost of fmaps accumulation.





\subsubsection{DRAM-Load-and-Store-related Attributes} \label{ch3.1.2:DLSA} DLSA includes \textbf{\textit{DRAM Tensor Order}} and \textbf{\textit{Living Duration}}, which will be introduced in serial.


The DRAM Tensor Order attribute determines the access sequence for all DRAM tensors, such as the tensors in dashed squares which is in the middle of Fig.~\ref{figure:example} Left.

Every DRAM tensor has an adjustable Living Duration attribute, which is a 2-element tuple $(Start,End)$. $Start$ and $End$ are tile IDs. This attribute has dual implications: first, within this time frame, the required buffer is allocated to this tensor. For the example in Fig.~\ref{figure:example}, the Living Duration of $W_{E}$ is $(B,D_{2})$. Therefore, $W_{E}$ lives in buffer from Tile $B$ to $D_{2}$. Second, it schedules the timing for loading and storing. Specifically, for ifmaps and weights, the $End$ is fixed at the next tile of the last tile that requires them, indicating when they can be released. The $Start$ indicates when this tensor can start being loaded from DRAM. For ofmaps, the $Start$ is fixed at the tile that produces them, indicating that once this tile finishes, storing to DRAM can begin. The $End$ signifies the tile by which the transfer must be completed; otherwise, the tile should stall. For example, the $End$ of $O_{E1}$ is $D_{1}$, so if its store is not completed, $D_{1}$ cannot start. Although $Start$ indicates when loading or storing can begin, the actual start time depends on whether the required data is ready and whether the loads and stores preceding this tensor in the DRAM Tensor Order have been completed. For example, the $Start$ of $I_{C2}$ is $C_{1}$, but the loading of the preceding $W_{D}$ takes a long time, causing the actual load of $I_{C2}$ to begin only in the middle of $E_{1}$.

\subsection{Space Size Comparison}\label{ch4:space_size}

\noindent As introduced above, each scheme is encoded by six variable attributes. Thus, our Tensor-centric Notation constructs a six-dimensional vast optimization space. In contrast, if we map the representable scheduling schemes of the SOTA Cocco~\cite{cocco} into our notation, only Computing Order and DRAM Cut can change, with the FLC Set being identical to the DRAM Cut Set. The other four attributes are determined by heuristic strategies, either explicitly or implicitly, making its explorable optimization space much smaller than ours. Moreover, DeFENIS~\cite{mei2023defines} proposes a simulator and demonstrates the performance and energy efficiency of: 1) different LGs with heuristically determined Tile Numbers, and 2) the same LGs with Tile Numbers enumerated at certain intervals. They analyze some characteristics of these two attributes individually. However, they do not jointly explore the space composed of these two attributes, let alone the other four dimensions. As a result, the schemes it touches are far fewer than ours.

 \begin{figure}[!t]

    \centering
    \includegraphics[width=3.2in]{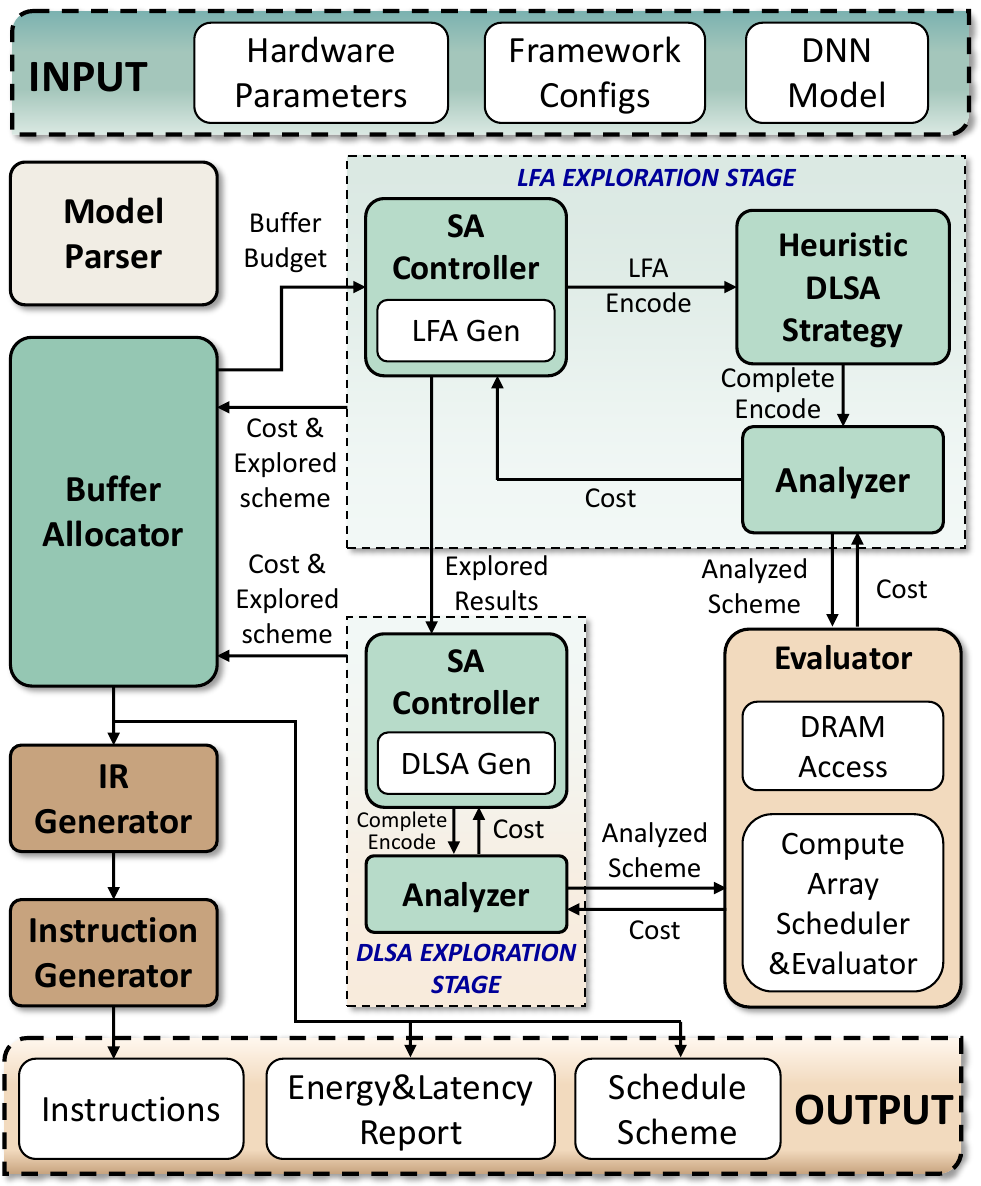}
    \caption{SoMa Framework}
    \label{figure:framework}
     \vspace{-6mm}
\end{figure}
\section{SoMa Framework}\label{sec:soma_framework}

\subsection{SoMa Overview}\label{sec:overview}
\noindent As shown in Fig.~\ref{figure:framework}, SoMa is an end-to-end DNN scheduling framework. SoMa takes as inputs: 1) hardware configuration, such as the number and organization of cores, DRAM bandwidth, buffer size, etc.; 2) framework configurations, such as the optimization goal and searching hyperparameters, etc.; 3) a DNN model description file generated by high-level frameworks like PyTorch After scheduling, SoMa outputs: 1) instructions; 2) reports on energy costs and latency; 3) a detailed scheduling scheme.


The optimization objective of SoMa is $Energy^n \times Delay^m$, where $n$ and $m$ are adjustable parameters that prioritize energy efficiency or performance according to different needs. $Energy$ and $Delay$ refer to the total energy costs and latency for processing a batch of samples, respectively. Thus, small batch sizes can evaluate SoMa's effectiveness in latency-centric scenarios, while larger batch sizes can be used for throughput-centric scenarios.

Fig.~\ref{figure:framework} shows that the key exploration process is governed by three main components: the Buffer Allocator, the LFA Exploration Stage, and the DLSA Exploration Stage. The Buffer Allocator controls the outermost iteration, each involving a complete two-stage top-down exploration. Based on the respective effects and buffer usage of the two stages in the previous iteration, the Buffer Allocator adjusts the buffer budget for the two stages competing for buffer usage in the current iteration. The two-stage top-down exploration involves separately varying and searching LFA and DLSA in each stage, with each stage using SA for independent searches. 

The reason for using a two-stage search with a Buffer Allocator is that: 1) as analyzed in Sec.~\ref{ch3_3:combine_them_together}, a single change in LFA can significantly impact DRAM tensors (e.g., adding or removing a DRAM Cut or changing the Tiling Number), thus previously optimized DLSA attributes may become sub-optimal or even invalid. This makes it difficult to retain good solutions obtained by varying DLSA attributes while continuously varying LFA. 2) The two-stage exploration with a Buffer Allocator can produce strong synergistic effects. DRAM access has a significant impact on performance and energy costs. Therefore, we observe that the first stage's optimization tends to minimize fmaps-related DRAM access, which reduces the optimization difficulty and increases the potential for the second stage optimization. This is mainly because a) the large proportion of weights is beneficial for optimization, as weights have fewer dependency constraints and can be adjusted more freely, while fmaps have more complex dependencies with a narrower range of adjustments; b) more extensive layer fusion and reduced DRAM access result in purer compute time, providing greater optimization opportunities for the second stage. Therefore, we found that dividing the optimization into two stages can produce good synergistic effects. The only risk is that the first stage might occupy too much buffer space, excessively limiting the optimization space for the second stage. We address this potential risk through the Buffer Allocator.

The optimal results can be directly output from the Buffer Allocator as scheduling schemes and reports, or they can be converted into a more easily parsable intermediate representation (IR) through our IR Generator module. This IR can then be fed into the Instruction Generator module to generate actual instructions.

\vspace{-2mm}
\subsection{Buffer Allocator}
\noindent In the first iteration, we conduct a complete two-stage search, with the only constraint being that the buffer usage does not exceed the hardware buffer capacity. We record the maximum buffer usage of the scheme explored in the first stage ($Buffer_{max}$), as well as the best overall encoding scheme and its cost ($Cost_{best}$). In subsequent iterations, the buffer usage limit for the first stage is reduced by $a\%$ (10\% used in the following experiments) of $Buffer_{max}$ each time (solutions exceeding this limit are deemed invalid), and the overall cost ($Cost_{temp}$) is recorded. If $Cost_{temp}$ is better than $Cost_{best}$, $Cost_{best}$ and the optimal encoding scheme are updated accordingly. The iteration stops when the costs of the optimal solutions found in two consecutive iterations do not exceed $Cost_{best}$. 
\SA{The rationale behind using this iteration to allocate buffers between the two stages for overall optimization is that while the performance of both stages improves with increased buffer size, the rate of improvement slows as buffer size grows. Therefore, adjusting the buffer allocation in small increments helps effectively find the sweet spot that maximizes the combined performance of both stages.}



\subsection{LFA \& DLSA Exploration Stage}

\noindent Both LFA and DLSA employ SA to explore this space. \SA{The key factors in SA are the initial solution, cooling schedule, and operators. The initial solution and operators are discussed in the following sections, while the cooling schedule is described here. Starting from an initial solution, each iteration randomly selects an operation to modify the encoding and evaluates it. If the new scheme’s cost ($c'$) is higher than the previous cost ($c$), it is accepted with probability $p = e^{\frac{c - c'}{c T_n}}$, where $T_n$ is the temperature at iteration $n$. Otherwise, the scheme is always accepted. The temperature at iteration $n$ is given by $T_n = T_0\frac{1-\frac{n}{N}}{1+\alpha\frac{n}{N}}$, where $T_0$ is the initial temperature and $\alpha$ is the cooling rate. The total number of iterations is $N = \beta X$. For the first stage, $\beta$ and $X$ are set to 100 and the number of layers, respectively. For the second stage, they are set to 1000 and the number of DRAM tensors, respectively. We also support setting an additional termination time. Once this time is reached, the algorithm performs $Y$ more iterations, accepting only improved solutions.}

\subsubsection{LFA Exploration Stage} \SA{In this stage, The initial solution consists of each layer forming its own independent LG and FLG (e.g., both $FLG$ and $LG$ are ${1, 2, 3, 4}$ as shown in Fig.~\ref{figure:example}), meaning no fusion is applied. The tile number is set to the minimum granularity, corresponding to the size required for the core array to perform parallel computation.} Then, the SA operators transform the LFA, while the DLSA is determined using a classical double-buffer strategy (as introduced in Sec.~\ref{ch3_3:combine_them_together}). The specific operators are as follows:

\noindent \textbf{Change Computing Order}: Randomly select a layer and change its order to another valid location.

\noindent \textbf{Change Tiling Number}: Randomly select an FLG and multiply or divide its Tiling Number by 2.

\noindent \textbf{Add/Delete An FLC}: Randomly add or delete an element in FLC Set. Specifically, adding an FLC means cutting an FLG into two FLGs with the same Tiling Number attribute as the original FLG. Removing an FLC means merging two FLGs into one, with the new FLG's Tiling Number inherited probabilistically based on the layer count ratio of the original two FLGs.

\noindent \textbf{Add/Delete A DRAM Cut}: Randomly add or delete an element in the DRAM Cut Set. The added element must be in the FLC Set.

\subsubsection{DLSA Exploration Stage}
\SA{In this stage, the initial solution adopts the best scheme explored by the previous stage, with the LFA attribute remaining constant.} The SA controller then primarily focuses on searching within the DLSA for the DRAM tensors corresponding to this selected LFA. The specific operators are introduced as follows:

\noindent \textbf{Change DRAM Tensor Order}: Randomly select a DRAM tensor and change its order to another valid location.

\noindent \textbf{Change Living Duration}: Randomly select a DRAM tensor and randomly change its $Start$ (for ifmaps and weights) or $End$ (for ofmaps). \SA{For example, in Fig.~\ref{figure:example}(b), by reducing the $Start$ of $W_B$ by 1 from $B$ to $A_2$, the STALL between $A_2$ and $B$ can be eliminated, and $W_B$ will be included in the buffer associated with $A_2$.}

\SA{Notably, in each operation, the probability of selecting a DRAM tensor is proportional to its size since larger tensors generally have a greater impact on performance and buffer utilization, warranting more transformation opportunities.}
\vspace{-1mm}
\subsection{Evaluator}\label{sec:evaluator}
\noindent In this section, we introduce an accurate evaluator, capable of evaluating various scheduling schemes, described using our Tensor-centric Notation across different hardware configurations in terms of energy cost and latency.

The evaluation process follows a local-to-global approach, first assessing each computing tile and DRAM load/store request (DRAM tensor) individually and then conducting an overall assessment.


\tool{For each computing tile (e.g., $A_{1}$ in Fig.~\ref{figure:example}), the ifmaps and weights have been prefetched into the GBUF, and the ofmaps are written back to the GBUF. From a macro perspective, the Core Array Scheduler explores how to further divide this tile into sub-tiles for computation by each core (as introduced in Sec.~\ref{ch2_hardware_baseline}), aiming to maximize parallelism and data reuse. 
The corresponding Evaluator assesses each interaction between GBUF and L0 buffers, as well as computations, while accounting for dependencies to evaluate overall performance and energy consumption. The corresponding energy costs and computing time of the searched optimal scheme are taken as the tile's energy costs and computing time. As this area of research is well-established~\cite{magnet,mindmapping,timeloop,understand,gammatushka,tenet,interstellar,CoSA,hasco}, we adopt a classic scheduler and evaluator for this purpose~\cite{understand,timeloop}.}


Each DRAM communication tensor's energy costs are calculated by summing the products of read and write data volumes with their respective unit energy costs. The read and write times are calculated by dividing the data volumes by the respective bandwidths.

The total energy cost is calculated by summing up the energy costs of the above sub-components, similar to existing classical works~\cite{understand,timeloop,interstellar}. The total computing time is derived based on the evaluated times of all computing tiles and DRAM tensors using the following method:

For each DRAM tensor, it can start execution only when the following three conditions are met: 1) the preceding DRAM tensor has been completed; 2) for ifmaps or weights, their $Start$ must be smaller than or equal to the current tile ID; and 3) for ofmaps, it must wait until its generating computing tile has finished. For example, in Fig.~\ref{figure:example}, although $I_{C2}$'s $Start$ is $C_{1}$, the preceding DRAM tensor ($W_{D}$) is not completed until $E_{1}$, so it can only start at the middle of $E_{1}$. Moreover, $W_{B}$ has a $Start$ of $B$. Although the previous DRAM tensor ($I_{A2}$) has already been completed, it must still wait until $A_{2}$ finishes before it can begin. 

Each computing tile can start execution only when the following conditions are met: 1) all required data (ifmaps, weights, etc.) are ready. For the example in Fig.~\ref{figure:motivation}, $A_{1}$, $B_{1}$, and $C_{1}$ cannot follow their respective preceding computing tile immediately because the required data are not ready at the end of the preceding computing tile. 2) All DRAM tensors with $End$ less than or equal to the tile must be completed. For the instance in Fig.~\ref{figure:example}, $D_{1}$ cannot follow $E_{1}$ because $O_{E1}$'s $End$ is $D_{1}$, and $D_{1}$ cannot start until $O_{E1}$ has finished execution.


\begin{figure*}[!t]
    \centering
    \includegraphics[width=7.2in]{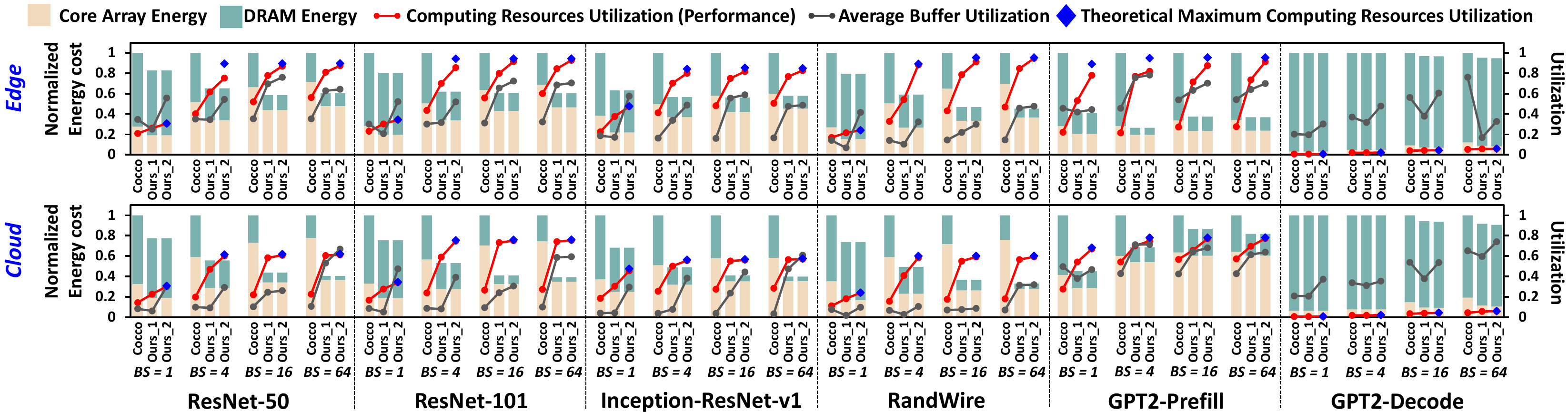}
    \caption{Overall Comparisons between Cocco and SoMa (Ours). Ours\_1 and Ours\_2 demonstrate how the efficient solution is progressively optimized through the first and second stages. 
    \textbf{Computing Resources Utilization} equals to $Util(\text{Actual Evaluated Latency})$, where $Util(t) = {\text{Total Number of Operations in Network}} / {(\text{Peak Hardware Computing Power} \times \text{t})}$. 
    Thus, \textit{the utilization can be viewed as a measure of \textbf{performance}}. \textbf{Average Buffer Usage} is calculated as $\sum_{Tile} (\text{Buffer Usage} \times \text{Computing\_Time}_{Tile}) / \text{Total Comp Time}$. The \textbf{Theoretical Maximum Computing Resources Utilization} (represented by \textcolor{blue}{Blue Diamonds}) is used to measure the theoretical maximum optimization possible in the second stage. Specifically, it is $Util(MIN\{\sum_{Tile} \text{Computing\_Time}_{Tile}, \sum_{Tensor} \text{DRAM\_Access\_Time}_{Tensor}\})$,
     without considering dependencies (e.g., $Util(MIN\{(I_{A1} + ... + O_{C4}),(A_1+ ... + C_4)\})$ in Fig.~\ref{figure:motivation}). This means the theoretical optimization upper limit in the second stage is achieved when all DRAM tensors or computing tiles are continuously executed without stall. BS represents batch size. For the edge platform, the workload is GPT-2-Small, while for the cloud platform, it is GPT-2-XL.}
    \label{figure:overall}
    \vspace{-2mm}
\end{figure*}

\subsection{Generality and Portability of SoMa}
\noindent The proposed encoding and SoMa framework exhibit excellent generality for two main reasons: 1) the template depicted in Fig.~\ref{figure:hardware} is highly general, encompassing many accelerators from both the industry~\cite{NNP-I,TPU,TPUlesson,huawei,cambriconmlu290} and academia~\cite{dadiannao,neurocube,tetris}; 2) The hardware behavior information encoded by our notation is very general (i.e., computing, data loading, and storing, instruction dependency, etc.).

Our SoMa also possesses excellent portability, due not only to the aforementioned reasons but also because our framework is designed with a robust modular architecture. This design allows for easy adaptation to different accelerators, which may have distinct core micro-architectures, by simply replacing the relevant Core Array Scheduler \& Evaluator modules and Instruction Generation module. We have developed a comprehensive compilation flow for our accelerator~\cite{Opensource}, which can serve as a concrete example for porting to other accelerators.

\subsection{Open-source of SoMa}\label{open}
\noindent\open{We have now open-sourced files and documentation for key stages to illustrate the entire SoMa Compiler workflow at this link~\cite{Opensource}. In the future, we plan to set up a small-scale cloud platform. This platform will allow users not only to access the open-sourced SoMa scheduler but also to modify or even replace our scheduler (as long as the output is converted to IR format), enabling it to be translated into instructions that can run on the chip. Given that many existing accelerators (e.g., TPU~\cite{TPU}) do not offer such low-level API access, we believe this platform can significantly advance related research.}
 
\section{Evaluation} 

\subsection{Experiment Setup}\label{sec:setup}
\subsubsection{Hardware Configuration}
\noindent \hardwareconfig{To comprehensively evaluate the effects of SoMa, we utilize both edge and cloud hardware platforms: we set the edge computing power to 16 TOPS (referencing Qualcomm Snapdragon 8 Gen 3's 15 TOPS~\cite{8gen3} and Apple A15's 15.8 TOPS~\cite{A15}, A16's 17 TOPS~\cite{A16}), and the cloud to 128 TOPS (referencing NVIDIA Orin's 138 TOPS~\cite{Orin} and TPU V4i's 136 TOPS~\cite{TPUlesson}). Buffer and DRAM bandwidth are set to 8MB and 32MB, and 16GB/s and 128GB/s, respectively, based on our DSE results in Fig.~\ref{figure:DSE}.} This configuration allows accelerators to achieve outstanding performance with reasonable hardware resources. We also thoroughly tested SoMa's performance under different buffer sizes and DRAM bandwidths (as shown in Fig.~\ref{figure:DSE}). The default process technology is TSMC 12nm, with an operating frequency of 1GHz. All unit energy costs for different operations required by the Evaluator (as introduced in Sec.~\ref{sec:evaluator}) are obtained from the actual RTL code synthesis and simulation during our accelerator development. The optimization goal is set as $\text{Energy}^1 \times \text{Delay}^1$, as both energy costs and performance are crucial in inference accelerators~\cite{meta_mtia,intel_inference_whitepaper}.

\subsubsection{Workload}\label{ch6_1:workload}
To comprehensively evaluate SoMa effects and analyze the trade-offs behind different DRAM communication scheduling schemes, we scale batch size from 1 to \experiment{64}, covering latency-sensitive scenarios to throughput-centric scenarios as introduced in Sec.~\ref{sec:overview}. In our experiments, ResNet-50~\cite{Resnet}, ResNet-101~\cite{Resnet}, Inception-ResNet-v1 (IRes)~\cite{inception-v4}, \experiment{Randwire~\cite{randwire} and GPT-2~\cite{GPT-2}} are chosen as workloads. ResNet-50 is chosen because it takes classical residual structures widely employed in many DNNs. ResNet-101 is chosen because it has a similar structure to ResNet-50 but with a larger number of layers. The Inception-ResNet-v1 and \experiment{Randwire} are selected to represent the DNNs with wider and more complex structures. \experiment{GPT-2 is selected to represent language-processing DNNs. Since GPT-2 has various versions for different scenarios, we use GPT-2-Small with a token length of 512 (prefill 512 and decode the 513th) for the edge platform, and GPT-2-XL with a token length of 1024 (prefill 1024 and decode the 1025th) for the cloud platform.}

ResNet-50 and \experiment{GPT-2} are chosen as the default workloads for the discussion due to their representativeness and popularity in image and language processing, respectively.

\subsubsection{Baseline}
\noindent \baseline{We select the SOTA Cocco scheduling framework~\cite{cocco} as our baseline, as it explores the layer-fusion space while adopting mainstream tiling and prefetch-and-delayed-storing strategies.}


\subsection{Overall Comparisons}\label{ch6_2:overall}

\noindent Fig.~\ref{figure:overall} shows that after the first stage, SoMa, on average, improves 1.82$\times$ performance and reduces 37.3\% energy costs compared to Cocco. The second stage, on average, further improves performance by 1.16$\times$ over the first stage. It can be observed that the schemes found in the second stage are very close to the theoretical maximum optimization value (blue diamonds), with an average difference of only 3.1\%. Thus, after optimization through the two stages, the best-explored solution improves 2.11$\times$ performance and reduces 37.3\% energy costs.


SoMa improves performance by 2.15$\times$, 2.18$\times$, 2.01$\times$, 2.61$\times$, 2.55$\times$ and 1.14$\times$ and reduces energy cost by 39.5\%, 41.0\%, 46.0\%, 47.1\%, 47.0\%, and 3.1\% on ResNet-50, ResNet-101, Inception-ResNet-v1, RandWire, GPT2-Prefill, and GPT2-Decode compared to Cocco, respectively. This demonstrates that SoMa can consistently improve performance and energy efficiency on various DNNs.
\textbf{However, compared to other networks, including GPT-2-Prefill, SoMa demonstrates almost no optimization effect in GPT-2-Decode, and the overall computational utilization remains extremely low. This is because the decode stage exhibits significantly lower compute density~\cite{splitwise}, with its latency primarily dominated by weight and KV cache loading. Moreover, another interesting phenomenon is that the computational utilization of GPT-2 does not increase linearly with the batch size; instead, the growth rate gradually diminishes. For instance, after SoMa optimization, the utilization rates for GPT-2-Small-Decode and GPT-2-XL-Decode across batch sizes of 1, 4, 16, and 64 are 0.66\%, 2.03\%, 4.26\%, 5.84\% and 0.60\%, 1.90\%, 4.13\%, 5.83\%, respectively. This is because, as the batch size grows, the increasing size of the KV cache becomes comparable to or even exceeds that of the weights, diminishing the benefits of further increases in batch size for improving compute density.}





\begin{figure*}[!t]
    \centering
    \includegraphics[width=7.2in]{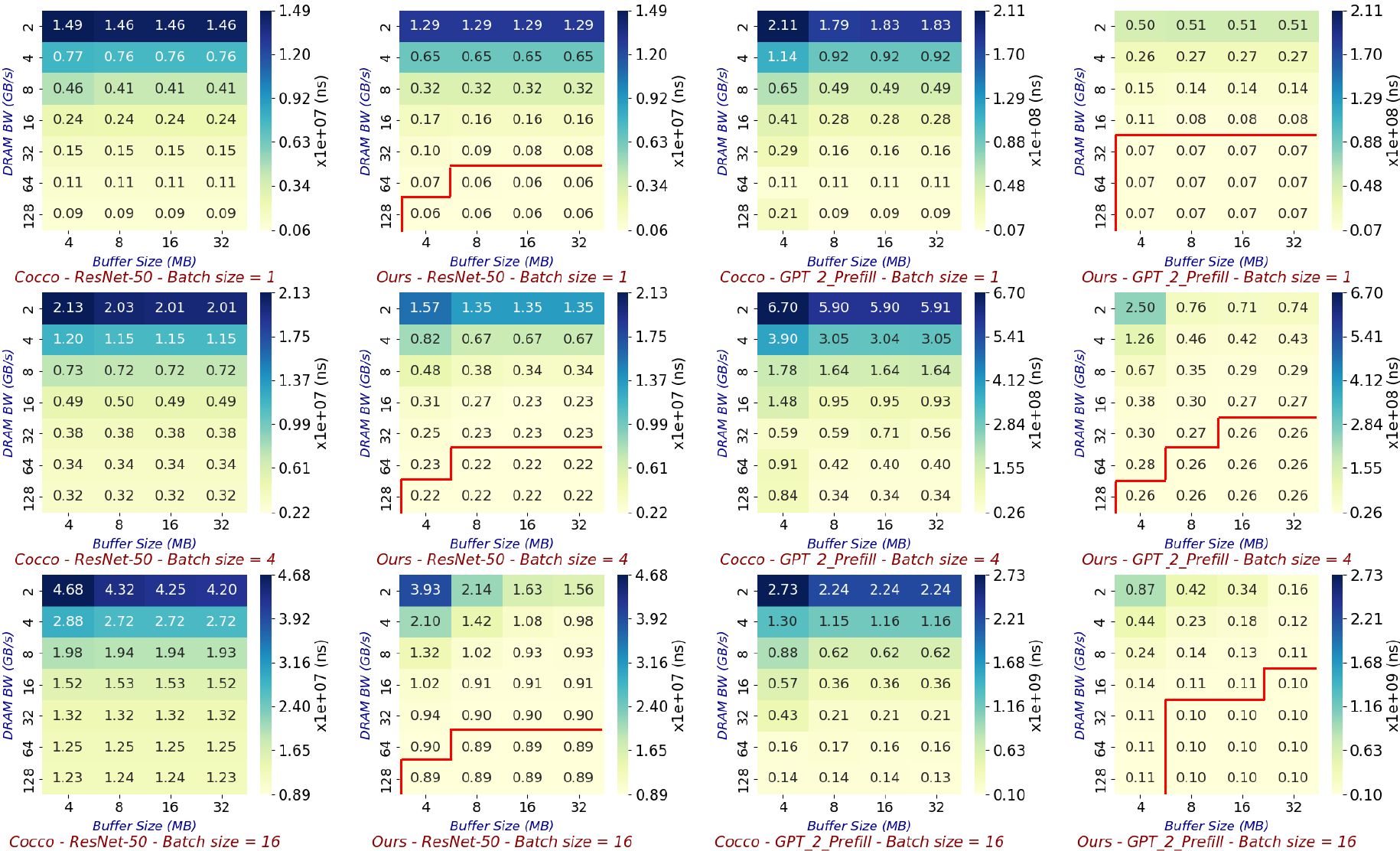}
    \caption{\experiment{Design Space Exploration over DRAM Bandwidth and Buffer Size for the 16TOPS Edge DNN Accelerator. We present the latency corresponding to the optimal schemes explored using Cocco and SoMa (Ours) under different networks, batch sizes, DRAM bandwidth, and buffer sizes. In the results of Ours, we highlighted the hardware configurations with the same minimum value (accurate to two decimal places) using a red envelope curve.}}
    \label{figure:DSE}
     \vspace{-2mm}
\end{figure*}

\subsubsection{Analysis of the First Stage of SoMa}
In the first stage of SoMa, we observe significant reductions in Core Array Energy and DRAM Energy by 34.8\% and 44.3\%, respectively. This reduction in Core Array Energy is mainly due to the flexibility of our approach in adjusting the Tiling Number during exploration, unlike Cocco's more conservative approach that selects each tile size based only on the basic parallelism requirements of the computing units~\cite{cocco} (detailed analysis and example can be found in Sec.~\ref{ch7:learn_1st}). Therefore, our approach often has fewer tiles within the buffer constraint (with an average total number of computing tiles per network being 7962 for Cocco and 751 for ours), \reuse{with each tile being larger, allowing more optimization and reuse opportunities in the Core Array Scheduler. The decrease in DRAM Energy is because SoMa allows for a smaller number of LGs per network (an average of 2.5) compared to Cocco (an average of 13.0), meaning it can fuse more layers.} The main reasons for this difference are: 1) Cocco's finer-grained tiles result in significant accumulated backtracking halo overlap costs when dealing with convolutional and pooling layers (as introduced in Sec.~\ref{ch2:fusion} and Fig.~\ref{figure:motivation}). 2) Our approach can use FLCs (with an average of 3.9 FLGs per network) instead of DRAM Cuts to free some weights (as introduced in Sec.~\ref{ch3.1.1:LFA}) and adjust the Tiling Number. Specifically, shuffling weights can save buffer space, enabling the fusion of more layers. Moreover, since different FLGs can have different Tiling Numbers, the FLC can switch the computing granularity between two FLGs without accessing DRAM (detailed analysis in Sec.~\ref{ch7:learn_1st}). 


The performance improvement also largely stems from the above-analyzed factors, as coarser-grained tiles provide more on-chip reuse opportunities, better overlapping of computation, and GBUF access time, while Cocco may sometimes suffer from GBUF load and store stalls. Additionally, the significant reduction in DRAM accesses naturally reduces overall latency.

\subsubsection{Analysis of the Second Stage of SoMa}

The performance improvement brought by the second stage mainly comes from effectively exploiting the DRAM usage imbalance opportunities described in Sec.~\ref{ch2_prefetch_delay}. Specifically, by intelligently prefetching and delaying storing, many DRAM idle periods can be utilized, reducing computation stalls and DRAM bandwidth waste. In Sec.~\ref{ch7:learn_2st}, we demonstrate this optimization through a simple practical example.
\begin{figure*}[!t]
    \centering
    \includegraphics[width=7in]{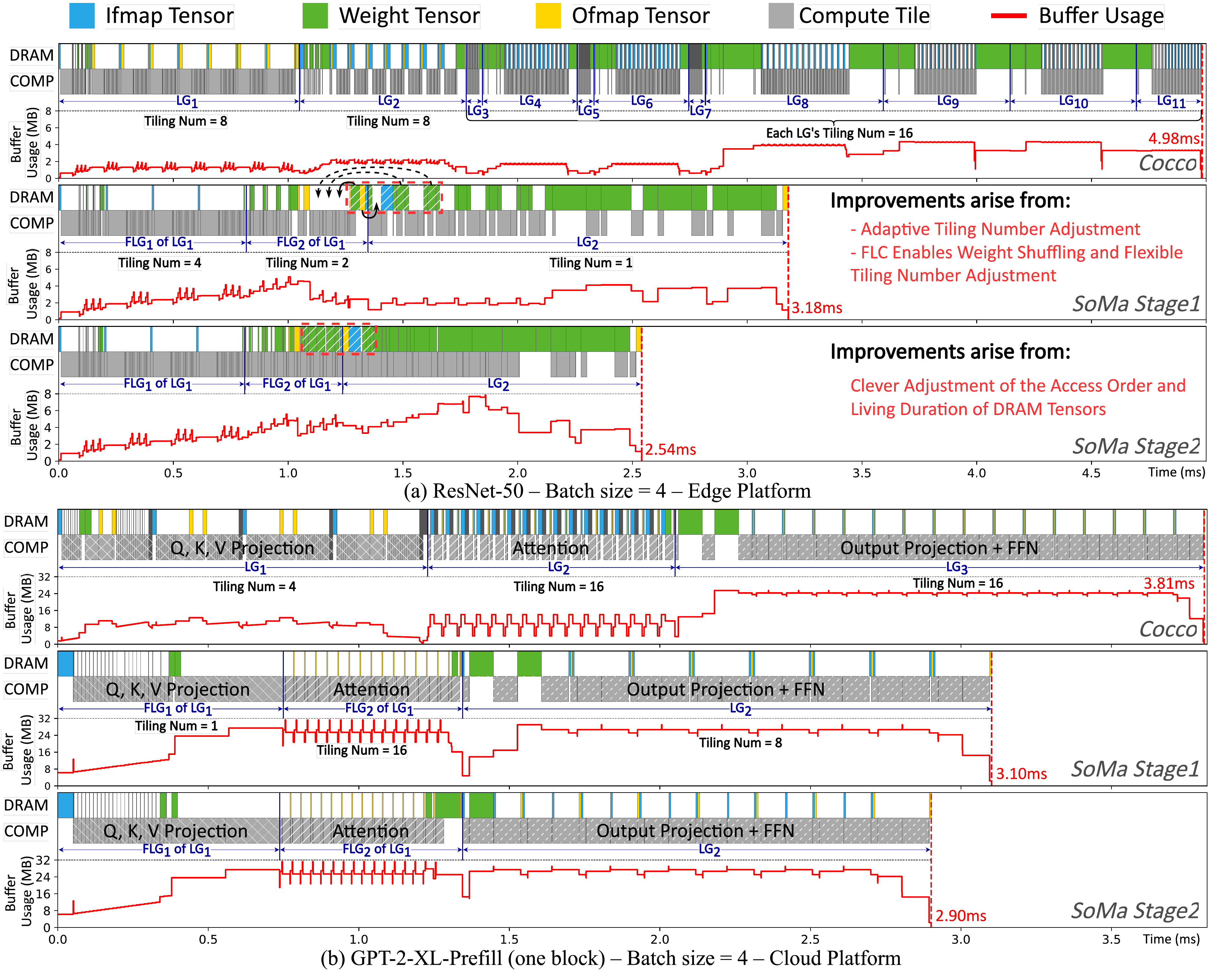}
    \caption{\experiment{Comparison of Practical Execution Graphs among the Schemes Explored by Cocco (top), the First Stage (middle), and the Second Stage (bottom) of SoMa on default edge-side accelerator. We also point out all DRAM Cuts, FLCs, and their corresponding Tiling Numbers. In the running graph for one block of GPT-2-XL-Prefill, we have highlighted the main matrix multiplication layers, while certain element-wise layers (such as transpose, softmax, add, and layer normalization) are not explicitly marked. Q, K, and V represents Query, Key, and Value, respectively. }}
    \label{figure:running_example}
    \vspace{-2mm}
\end{figure*}

\section{Discussion}

\subsection{Design Space Exploration}
In Fig.~\ref{figure:DSE}, we conduct a DSE for a 16 TOPS accelerator across different workloads and batch sizes, focusing on buffer size and DRAM bandwidth. This exploration yields some interesting insights as follows.

\subsubsection{Insight 1}
\textbf{When dealing with small batch sizes (especially 1), even with SoMa, DRAM bandwidth plays a more decisive role in performance compared to buffer size. However, the impact of buffer size becomes more significant as the batch size increases.} 

For small batch sizes (e.g., Ours-Batch size = 1 in Fig.~\ref{figure:DSE}), the greater influence of DRAM bandwidth is evident as increasing buffer size does not significantly reduce latency, even with SoMa. However, increasing DRAM bandwidth results in a noticeable reduction in latency. This is because: 1) the fmaps sizes are small, so a minimal buffer can almost entirely store data on-chip, making further increases in buffer size unnecessary; 2) weights need to be loaded regardless, but a small batch size means less computation, providing fewer opportunities for SoMa to use prefetching and delayed storing to hide data transfer times. Consequently, weight loading times can severely stall computation and dominate latency. Thus, increasing DRAM bandwidth can directly reduce this latency and significantly enhance overall performance.

As the batch size increases, the impact of buffer size becomes more significant, as evidenced by the greater reduction in latency with larger batch sizes when increasing buffer size (e.g., Ours-Batch size = 4 and 16 in Fig.~\ref{figure:DSE}). This is because: 1) the fmaps sizes increase, allowing a larger buffer to better exploit reuse opportunities; 2) additionally, with more computation, the longer computing times provide more opportunities for SoMa to use the buffer for prefetching and delayed storing to hide data transfer times.

\subsubsection{Insight 2} \textbf{Having both a large buffer and high DRAM bandwidth is often wasteful and offers limited performance gains compared to configurations with either high DRAM bandwidth or moderate DRAM bandwidth paired with a large buffer.}

In Ours in Fig.~\ref{figure:DSE}, we observe that many configurations in the lower right corner exhibit similar performance (highlighted by a red envelope curve). Among these points, the ones in the farthest lower right corner are equipped with both the large DRAM bandwidth and buffer sizes, which is quite wasteful. Specifically, we find that for all workloads, there is a noticeable point where increasing DRAM bandwidth beyond a certain threshold yields diminishing returns, as the dominant factor for latency becomes computing time. This lower triangle (red envelope curve) in SoMa, a feature not observed in Cocco, indicates that with the efficient buffer utilization in SoMa, on-chip buffers can adequately compensate for DRAM bandwidth. While HBM provides high bandwidth density in the current era, its high cost deters many companies~\cite{tenstorrent,DRAMPRICE}. Therefore, moderately increasing buffer size and opting for lower-cost DRAM models could be a viable differentiation strategy, and SoMa can be a significant aid for this.


\subsection{Lessons From a Practical Example}

\noindent Fig.~\ref{figure:running_example} shows the actual execution graph of scheduling schemes explored by Cocco, SoMa's first stage, and SoMa's second stage, respectively. For ResNet-50, compared to the baseline, the first stage of SoMa achieves an average of 1.57$\times$ performance improvement and a 36.1\% reduction in energy cost. The second stage achieves an additional average of 1.25$\times$ performance improvement over Stage 1, resulting in a total performance gain of 1.96$\times$ over Cocco. Next, we will analyze the reasons for SoMa's performance and energy efficiency improvements using this practical example, thereby enhancing the understanding of the DRAM Communication Scheduling Space.

\subsubsection{Lessons From the First Stage}\label{ch7:learn_1st} From this example, we can make two observations: 1) \textbf{adaptively adjusting the Tiling Number based on network characteristics and buffer size is crucial for exploiting buffer potential and improving performance; 2) FLC can free weights and aggregate fmaps to adjust the Tiling Number without accessing DDR, which can further enhance buffer utilization efficiency and leverage the potential of dynamically adjusting the Tiling Number.} Next, we will analyze the origins of these two insightful observations in detail, using the example in Fig.~\ref{figure:running_example} as a reference.

First, by comparing the Tiling Numbers in Cocco and the first stage of SoMa, we find that Cocco's Tiling Number is larger, while SoMa’s Tiling Number can be as low as 1. This is mainly because Cocco uses a heuristic strategy to set the Tiling Number based on the Core Array's parallelism requirements (Kernel-Channel (KC) parallelism~\cite{TPU,huawei,nvdla,tpuv2,TPUlesson,tenstorrent}), a common approach in many studies~\cite{atomic,EfficientSchedulingliuleibo}. Consequently, larger kernel and channel dimensions result in a higher Tiling Number. For example, in ResNet-50, the Tiling Number (16) for later layers with larger kernel and channel dimensions is higher than that of earlier layers (8). However, this strategy is often conservative, leading to higher Tiling Numbers. A larger Tiling Number increases backtracking halo overlap costs and limits the Core Array Scheduler's ability to find on-chip reuse opportunities, which reduces performance and energy efficiency. In contrast, SoMa better adapts to buffer size and network structure to control the Tiling Number. For example, in ResNet-50, the fmap size decreases as the network deepens, so the buffer consumption for a smaller Tiling Number also decreases accordingly. Thus, SoMa intelligently allows the Tiling Number to decrease monotonically with network depth (from left to right), rather than increasing as in Cocco. \experiment{Additionally, we observe that in SoMa, the last LG of ResNet-50 and the first FLG of GPT-2-XL-Prefill both have a Tiling Number of 1, allowing for the immediate disposal of weights after processing each tile (or layer). This conserves buffer space and facilitates further fusion.}

\experiment{Second, in both ResNet-50 and GPT-2-XL-Prefill, it happens that the first two LGs in the Cocco scheme and the first two FLGs in the SoMa scheme consist of identical layers. We observe that SoMa efficiently uses the FLC to clear weights and adjust tile sizes without incurring the DRAM access overheads seen in the Cocco scheme. Specifically, in ResNet-50, if the Tiling Number for the first FLG equals 2 (matching the next FLG), the buffer is insufficient for the first FLG. In GPT-2-XL-Prefill, only when the Tiling Number for the first FLG is 1 can it promptly clear weights to accommodate Q, K, and V on chip.}

\subsubsection{Lessons From the Second Stage}\label{ch7:learn_2st} The improvement of the second stage is attributed to the clever adjustment of the access order and living duration of DRAM tensors. In this stage, SoMa properly performs precise surgical strikes on some key tensors to reduce the total computing stall.

\experiment{It can be observed that in both examples, SoMa’s second stage effectively prefetches and delays storing DRAM tensors. ResNet-50, however, has more DRAM tensors, making adjustments more complex; therefore, we primarily use it as the example for analysis.}

As shown by the black solid arrows near the diagonally striped regions in Fig.~\ref{figure:running_example}, DRAM tensors that cause computing stalls are adjusted to DRAM-free periods. More specifically, the weight of the first layer of $LG_2$ is prefetched several tiles earlier, while the ofmaps of the last layer of $FLG_2$ is delayed by one tile and swaps positions with the ifmaps of the first layer of $LG_2$. This adjustment successfully utilizes DRAM idle time to eliminate the computing stall near the border between $FLG_2$ and $LG_2$, while considering dependencies.

Additionally, it can be observed that $LG_2$ contains many chunky weights (with the three largest weights up to 2304KB with INT8 precision). Given the limited 8MB buffer, prefetching these weights far in advance is impractical. However, simply pushing all weights forward would not maximally eliminate computing stalls. Therefore, SoMa cleverly selects to move the two relatively larger weight tensors at the front of $LG_2$ to the DRAM idle time at the back of $FLG_2$ (marked by dashed black arrows). The remaining weights are not altered in their DRAM Access Order but are pushed forward by changing their Living Duration. This approach maximizes the elimination of computing stalls within $LG_2$. Although some computing stalls still appear, the maximum buffer usage near these stalls has already reached 8MB, indicating that further prefetching is not feasible.


\section{Conclusion}

\noindent In this work, we first analyze various existing paradigms that use buffers to optimize DRAM communication, identifying challenges in the current layer fusion paradigm and uncovering previously overlooked opportunities for prefetching and delayed storing. Based on these observations, we introduce a Tensor-centric Notation and a matching parsing method to describe DRAM communication scheduling schemes, thereby defining the DRAM Communication Scheduling Space. We then propose a two-stage scheduling framework, SoMa, to structurally explore this space. Experimental results demonstrate that, compared to the SOTA Cocco framework, SoMa efficiently explores the broader optimization space defined by our notation, fully utilizing the buffer's potential to optimize DRAM communication. Moreover, we leverage SoMa to conduct several case studies, yielding interesting insights into architecture design and the trade-offs underlying the DRAM Communication Scheduling Space.
\section{Acknowledgment}
\noindent This research was partially supported by Dushi Program from Tsinghua University.

\bibliographystyle{IEEEtranS}
\bibliography{main}
%
%
%
%
%


\appendix
\section{Artifact Appendix}

\subsection{Abstract}

\noindent This appendix provides guidance on accessing and using the SoMa framework (introduced in Sec.~\ref{sec:soma_framework}) to replicate the key results shown in Fig.~\ref{figure:overall} and Fig.~\ref{figure:DSE}. It also presents the architecture of the end-to-end compiler built on SoMa, detailing the inputs and outputs of each major stage (from model input to final instructions). Other experiments, which also involve similar steps and analyses, are omitted here for the sake of brevity.
\vspace{-1mm}

\subsection{Artifact check-list (meta-information)}


{\small
\begin{itemize}
  \item {\bf Algorithm: Simulated Annealing}
  \item {\bf Program: C++, Shell, Python (only for data collection)}
  \item {\bf Compilation: by Makefile}
  \item {\bf Hardware: Recommend a server with 96+ cores and at least 1GB RAM per core.}
  \item {\bf Metrics: Cost function $E\times D$ is employed in all experiments.}
  \item {\bf Experiments: reproduce Fig.~\ref{figure:DSE} and Fig.~\ref{figure:overall}.}
  \item {\bf How much disk space required (approximately)?: 1GB}
  \item {\bf How much time is needed to prepare workflow(approximately)?: Several minutes at most.}
  \item {\bf How much time is needed to complete experiments (approximately)?: For all 432 experiments (96 for Fig.~\ref{figure:overall} and 332 for Fig.~\ref{figure:DSE}), it takes about 2 days on a 192-core Intel Xeon Platinum 8260. Most experiments (95\%) are completed within 3.5 hours, while the remaining ones, mainly experiments with batch=64, require the full 2 days to finish.}
  \item {\bf Publicly available?: Yes}
  \item {\bf Code licenses (if publicly available)?: AGPL-3.0 License
}
  \item {\bf Archived (provide DOI)?: 10.5281/zenodo.14599935}
\end{itemize}
}

\subsection{Description}

\subsubsection{How to access}

The artifact is uploaded to Zenodo:10.5281/zenodo.14599935


\subsubsection{Software dependencies}

A C++ compilation environment with support for the C++ 17 standard is required. Linux is recommended. It is recommended to use ``GNU make'' to build the program. Additionally, the following Python packages are required for reproducing Fig.~\ref{figure:DSE}: ``pandas'', ``matplotlib'', ``seaborn'', and ``numpy''.



\subsection{Installation}

\noindent For artifact evaluation, start by downloading the artifact from Zenodo:

\lstset{basicstyle=\ttfamily\footnotesize,
  breaklines=true
}

\begin{lstlisting}
  $ wget -O SOMA_AE.zip https://zenodo.org/records/14599935/files/SOMA_AE.zip?download=1
  $ unzip SOMA_AE.zip
\end{lstlisting}

Our SoMa exploration framework is in ``SOMA''.
We use `build.sh` to build the SoMa framework and create the result directory.

\begin{lstlisting}[]
  $ cd SOMA
  $ ./build.sh
\end{lstlisting}
The executable target will be generated at ``./build/soma'', and the result directories will be ``results/overall'' and ``results/dse''.

You can install the needed Python packages using pip with the following commands.
\begin{lstlisting}[]
  $ pip install -r requirements.txt
\end{lstlisting}
Or you can use conda to install with the following commands.
\begin{lstlisting}[]
  $ conda install --file requirements.txt
\end{lstlisting}

\subsection{Experiment workflow}

\subsubsection{Overall and DSE}
Once the SoMa framework is built, you can reproduce the Overall (Fig.~\ref{figure:overall}) and DSE (Fig.~\ref{figure:DSE}) experiments with the command below.

\begin{lstlisting}[]
  $ ./run.sh --eta
\end{lstlisting}
The ``run.sh'' takes parameters from ``args.txt'' as input for each soma instance, including compute power, DRAM bandwidth, storage directory, and the random seed. For each configuration, both our method and the baseline use the same seed. By default, the ``run.sh'' utilizes all CPU cores, with each core running a separate SoMa process. Each SoMa process outputs the corresponding results and logs to ``results/overall'' and ``results/dse''. When using all 192 cores, it takes around 2 days to run on an Intel Xeon Platinum 8260 server. Due to the long runtime, we recommend using tools like ``nohup'' or ``screen'' to prevent disconnection due to inactivity, ensuring that all experiments complete successfully.

After all experiments are completed, we use ``get\_results.sh'' to extract the results from the raw outputs.

\begin{lstlisting}[]
  $ ./get_results.sh
\end{lstlisting}

``get\_results.sh'' will use Python scripts under folder ``pyscripts'' to generate four files: ``overall.csv'', ``stats.log'', ``dse.csv'', and ``Fig7\_heatmaps\_DSE.svg''. ``overall.csv'' contains all the data presented in Fig.~\ref{figure:overall}, while ``stats.log'' includes all the data analyzed and calculated in the Sec.~\ref{ch6_2:overall}. ``dse.csv'' contains all the data related to Fig.~\ref{figure:DSE}, and ``Fig7\_heatmaps\_DSE.svg'' reproduces Fig.~\ref{figure:DSE}.

\subsubsection{Comparison with Baselines} 
The comparison with the baseline in Fig.~\ref{figure:overall} can be found in ``stats.log''. For detailed data of each case, you can refer to ``overall.csv''. The results of the DSE experiment are available in ``Fig7\_heatmaps\_DSE.svg'', with the detailed data in ``dse.csv'', which is also human-readable, just like ``overall.csv''.

\subsubsection{SoMa Compiler Workflow}
In ``Compiler-IR'', we present open-sourced files that showcase the workflow of the end-to-end SoMa-based compiler developed for our high-performance commercial AI accelerator, the ZEBU FPGA-based Verification Platform, and the corresponding results. For related information, please refer to ``Compiler-IR/Readme.md''. (This section is for material demonstration only and does not require execution or reproduction.)

While we are currently unable to release the full source code of the whole compiler due to IP flow restrictions, we believe the provided scheduling engine at the core of SoMa, along with the materials and documentation, effectively offers a clear understanding of the entire workflow of the SoMa-based compiler. Additionally, we are committed to establishing a small-scale cloud platform after the chip tape-out and related testing are completed. This platform will allow users to access the open-sourced compiler based on SoMa, with the flexibility to modify or even replace our scheduler (as long as the output is converted into IR format), enabling translation into chip-executable instructions. (This commitment has also been included in Sec.~\ref{open} of the paper.)

\subsection{Evaluation and expected results}

\noindent After executing ``./get\_results.sh'', you can use the following script to compare the results with the expected ones:

\begin{lstlisting}[]
  $ compare.sh
\end{lstlisting}

``compare.sh'' uses the ``diff'' command to compare ``overall.csv'', ``stats.log'', and ``dse.csv''. The ``Fig7\_heatmaps\_DSE.svg'' file may not be byte-for-byte identical due to font or library version differences, but the data used, namely ``dse.csv'', should be identical. If all files match, it will output a message like ``All files match the expected results!''. Otherwise, it will report ``Some files do not match the expected results.'' and highlight the differences.



\subsection{Methodology}

Submission, reviewing and badging methodology:

\begin{itemize}
  \item \url{https://www.acm.org/publications/policies/artifact-review-badging}
  \item \url{http://cTuning.org/ae/submission-20201122.html}
  \item \url{http://cTuning.org/ae/reviewing-20201122.html}
\end{itemize}



\end{document}